\documentclass[reprint,amsmath, superscriptaddress, amssymb, aps,longbibliography,prl,showkeys,citeautoscript]{revtex4-1}

\usepackage{graphicx}
\usepackage{dcolumn}
\usepackage{bm}
\usepackage{nccmath}

\usepackage[T1]{fontenc}
\usepackage[utf8]{inputenc}
\usepackage{physics}
\usepackage{lineno}

\usepackage{xcolor}
\usepackage{pgffor}
\newcommand\hl[1]{%
  \bgroup
  \hskip0pt\color{blue}%
  #1%
  \egroup
}
\DeclareMathAlphabet{\pazocal}{OMS}{zplm}{m}{n}

\setcitestyle{super}

\begin{document}


\title{ARPES signatures of trions in van der Waals materials}

\author{Giuseppe Meneghini}
\email{giuseppe.meneghini@physik.uni-marburg.de}
\affiliation{Department of Physics, Philipps-Universit\"at Marburg, D-35032 Marburg, Germany}
\affiliation{mar.quest | Marburg Center for Quantum Materials and Sustainable Technologies,  Hans-Meerwein-Straße 6, D-35032 Marburg, Germany}

\author{Maja Löwe}
\affiliation{Department of Physics, Philipps-Universit\"at Marburg, D-35032 Marburg, Germany}
\affiliation{mar.quest | Marburg Center for Quantum Materials and Sustainable Technologies,  Hans-Meerwein-Straße 6, D-35032 Marburg, Germany}

\author{Raul Perea-Causin}
\affiliation{Department of Physics, Stockholm University, AlbaNova University Center, Stockholm, Sweden}

\author{Jan Philipp Bange}
\affiliation{%
 I. Physikalisches Institut, Georg-August-Universität Göttingen, Göttingen, Germany}%

 \author{Wiebke Bennecke}
\affiliation{%
 I. Physikalisches Institut, Georg-August-Universität Göttingen, Göttingen, Germany}%

\author{Marcel Reutzel}
\affiliation{Department of Physics, Philipps-Universit\"at Marburg, D-35032 Marburg, Germany}
\affiliation{mar.quest | Marburg Center for Quantum Materials and Sustainable Technologies,  Hans-Meerwein-Straße 6, D-35032 Marburg, Germany}

\author{Stefan Mathias}
\affiliation{%
 I. Physikalisches Institut, Georg-August-Universität Göttingen, Göttingen, Germany}%

\author{Ermin Malic}
\affiliation{Department of Physics, Philipps-Universit\"at Marburg, D-35032 Marburg, Germany}
\affiliation{mar.quest | Marburg Center for Quantum Materials and Sustainable Technologies,  Hans-Meerwein-Straße 6, D-35032 Marburg, Germany}

\date{\today}

\begin{abstract}
Angle-resolved photoemission spectroscopy (ARPES) has recently emerged as a direct probe of excitonic correlations in two-dimensional semiconductors, resolving their dispersion and dynamics in energy–momentum space, including dark exciton states inaccessible to optical techniques. However, the ARPES fingerprint of charged excitons (trions), which plays a key role in all doped and gated 2D material systems, has remained unknown so far. We present a first theoretical analysis of trion signatures in monolayer transition-metal dichalcogenides, highlighting how the additional charge carrier modifies the spectral position and shape relative to neutral excitons in ARPES spectra. Interestingly, we further predict that mass-imbalanced trions yield a characteristic double-peak structure, clearly separated in energy and line shape from neutral excitons. The predicted temperature dependence of these features offers guidance for experimental investigations aimed at identifying trionic states, thereby establishing a framework for ARPES studies of many-body Coulomb complexes in doped two-dimensional semiconductors.
\end{abstract}

\maketitle
Angle-resolved photoemission spectroscopy (ARPES) has become an indispensable tool for probing the electronic structure of semiconductors \cite{damascelli2003angle,sobota2021angle,boschini2024time,reutzel2024probing}. In two-dimensional materials, the reduced screening allows excitonic resonances to be directly observed \cite{christiansen2019theory,madeo2020directly,man2021experimental,wallauer2021momentum,dong2021direct,schmitt2022formation,bange2023ultrafast}. In contrast to optical techniques, ARPES provides simultaneous access to both photoelectron energy and momentum, enabling the direct detection of momentum-dark excitonic states \cite{madeo2020directly,schmitt2022formation,meneghini2023hybrid,bange2024probing,werner2025role, hagel2021exciton,bennecke2025hybrid,reutzel2024probing} that otherwise can appear only indirectly through typically weak phonon sidebands in photoluminescence spectra \cite{brem2020phonon,PhysRevResearch.1.032007,he2020valley,rosati2020temporal,funk2021spectral}. This ability to map the full Brillouin zone makes ARPES uniquely suited to investigate the relaxation dynamics of photoexcited quasiparticles, which govern the optical response and device performance of van der Waals heterostructures \cite{mueller2018exciton}. While ARPES signatures of neutral excitons in transition-metal dichalcogenides (TMDs) have been well established, much less is known about doped systems, where at low temperature and carrier density the response is dominated by charged excitons (trions) \cite{mak2013tightly,ross2013electrical,singh2016trion,plechinger2016trion,PhysRevB.96.085302,sidler2017fermi,arora2019excited,perea2022trion,perea2024trion}. 
\begin{figure}[t!]
  \centering
  \includegraphics[width=\columnwidth]{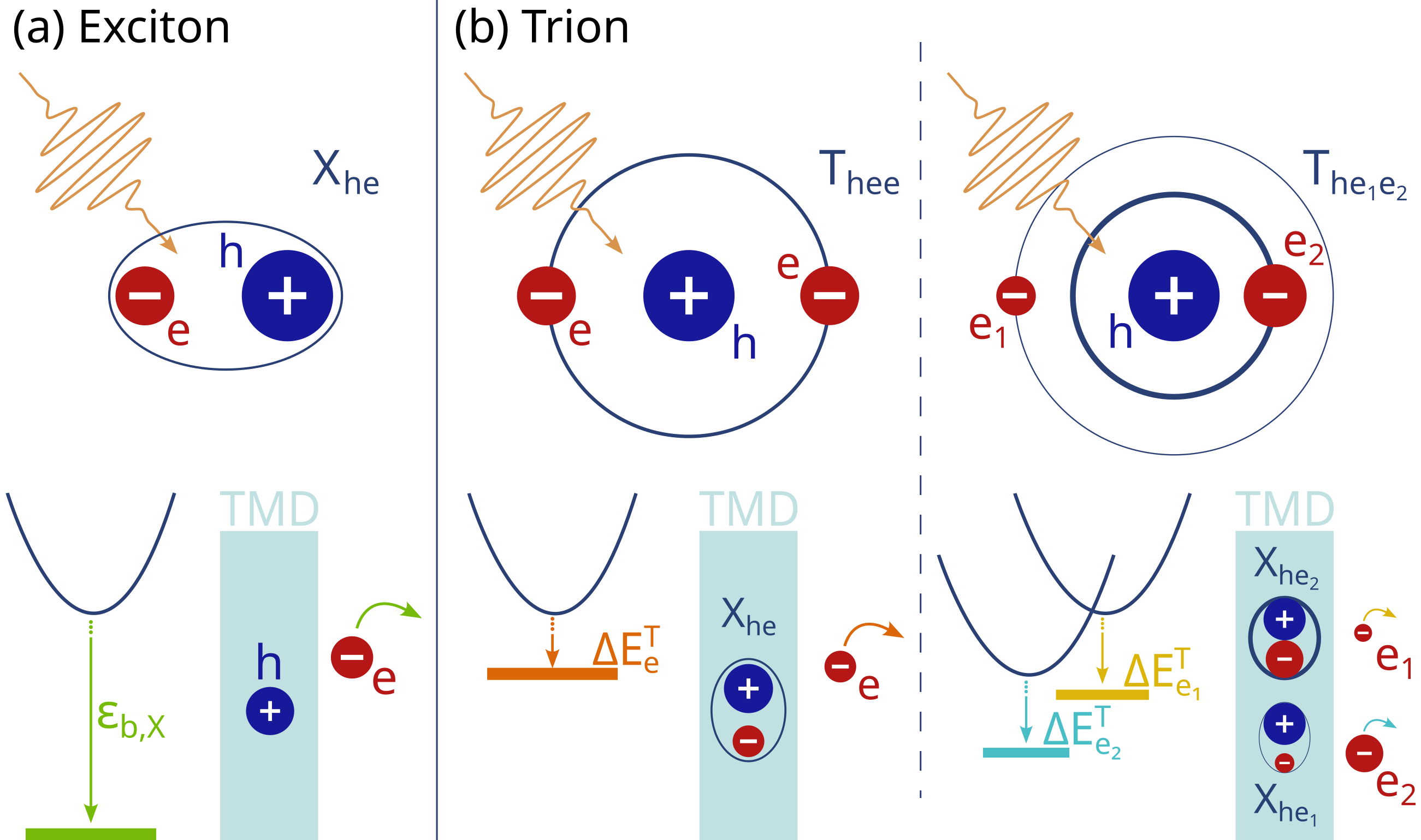}
  \caption{
Schematic illustration of exciton and trion contributions to the ARPES spectrum of an n-doped semiconductor. After exciton/trion formation, the photoemission pulse (orange wavy line) breaks the quasiparticle (electron–hole or electron–electron–hole complex), ejecting one electron and leaving behind a hole/exciton. Trions and excitons are denoted by T$_{\rm{h, e_1, e_2}}$ and X$_{\rm{h, e_{1/2}}}$, respectively, with the subindices describing the constituent electrons ($e_{1/2}$) and holes ($h$). The thickness of the lines representing electron orbits and excitons indicates the relative binding strength with thicker lines denoting a stronger binding.
(a) Exciton case: the photoemitted electron is ejected from the material and its signal in ARPES is located one exciton binding energy ($\varepsilon^{}_{b,X}$) below the conduction band minimum.
(b) Trion case: in the case of mass-balanced trions (left panel), the two electrons have a similar effective mass, so emission of either yields the same final configuration, producing a single spectral feature in ARPES located one electron-exciton binding energy ($\Delta$E$^T_e$) below the conduction band minimum. 
 In the case of mass-imbalanced trions (right panel), electrons with different effective masses have unequal binding energies to the residual exciton ($\Delta$E$_{e_{1/2}}^T$), leading to a characteristic double-peak structure in ARPES spectra.
}
   \label{fig:scheme}
\end{figure}

In this work, we address this knowledge gap through a microscopic and material-specific analysis of photoemission spectra of electrons bound to excitons in doped TMD monolayers. Taking WSe$_2$ as a representative case and focusing on n-type doping, we reveal the spectral position and shape of trion signatures in ARPES. We predict photoemission resonances arising from trions to be located one electron--exciton binding energy (tens of meV) below the conduction band minimum (see Fig. \ref{fig:scheme}b). Thus, the trion signal is clearly separated from the excitonic ARPES resonance, whose spectral position is lowered by the exciton binding energy (hundreds of meV, see Fig. \ref{fig:scheme}a). Furthermore, we show that the shape of the energy-momentum resolved trion signal is also markedly distinct from the excitonic shape that reflects the negative curvature of the valence band. Most interestingly, we predict a characteristic double-peak feature  arising from mass-imbalanced trions containing two electrons from different valleys (see Fig. \ref{fig:scheme}b, right panel), and analyze the temperature dependence. To the best of our knowledge, this work is the first to clearly reveal distinguishable trion features in ARPES. Our results and methodology establish a general framework for studying the photoemission fingerprints of many-body Coulomb complexes in doped two-dimensional semiconductors.

\emph{Microscopic model}---%
The ARPES intensity can be described within a time-dependent perturbation theory by the Fermi’s golden rule  \cite{damascelli2003angle,meneghini2023hybrid} 
\begin{align}
    \mathcal{I}({\bf k},h\nu; t) \propto \sum_{i f} \,\lvert \bra{f, \bf k} H_{int} \ket{i}\rvert^2 N_i(t)\,
    \delta\left(\Delta E_{f,i, \bf k} \right)
  \label{eq:arpes1}
\end{align}
with $\ket{i/f}$ and $E_{i/f}$ denoting the initial/final states and their corresponding energy with $N_i(t)$ as the occupation of the initial state. We discard the explicit time dependence of the intensity and the occupation, focusing on the static thermalized ARPES, arising from a Boltzmann occupation of the initial state. Furthermore,  $\Delta E_{f,i, \bf k}=E_{f,\textbf{k}}- E_i - h\nu$ describes the energy conservation during the photoemission process with the photon energy  $h\nu$ and with $E_{f,\textbf{k}}$ as the energy of the ejected  free electron with the momentum ${\bf k}$.   The interaction Hamiltonian $H_{int} = \sum_{\alpha\beta} \mathcal{M}_{\alpha\beta} a^{\dagger}_{\mathrm{f}\alpha} a^{}_{c\beta}$ describes the excitation of an electron from the conduction band to the free state with $a^{(\dagger)}$ denoting the electronic creation/annihilation operator. Note that we use the suffix $\mathrm{f}$ for a free state and $c/v$ for conduction/valence band states. The optical selection rules are contained in the optical matrix element $\mathcal{M}_{\alpha \beta}$ (see the Supplemental Material, SM, for more details). 

In this work, we focus on trion features in ARPES spectra, considering the exemplary case of an n-doped 2D material  (the formalism remains identical for p-type doping). The initial state is a trion, which is broken into its constituents by photoemission. In the first-order process considered here, one of the trion's electrons is photoemitted, leaving behind a bound exciton. The corresponding final state thus consists of a free electron and an exciton state. For comparison, in the excitonic photoemission signal, the final state consists of a single photoemitted conduction-band electron and an unbound hole in the valence band \cite{schmitt2022formation,damascelli2003angle,sobota2021angle,madeo2020directly,weinelt2004dynamics}. Note that in the trion case, a simultaneous ejection of both electrons or ejection of one electron leaving behind an unbound electron-hole pair could in principle also occur. In the latter case, we expect an exciton to be formed quickly due to the strong attractive Coulomb interaction. In the former case, we expect the resulting signal to be much weaker, as it requires a multi-photon process.   

To gain access to the exciton and trion energy landscape and wavefunctions, we start from the Hamilton operator in the electron-hole picture, including the attractive electron-hole Coulomb interaction in the case of excitons, and both electron-electron and electron-hole interactions in the case of  trions. Then, we solve the respective Wannier-like Schrödinger equations for two and three interacting charges \cite{PhysRevB.88.045318,brem2018exciton,PhysRevB.101.195417,perea2022trion,perea2024trion}
\begin{equation}
\begin{alignedat}{2}
H_{e,h} \text{ }\psi^\mu(\textbf{k})  &=  \varepsilon^\mu_{b,X}\text{ }\psi^\mu(\textbf{k})\\
    H_{e,e,h} \text{ }\phi^\eta(\textbf{k},\textbf{p})  &=  \varepsilon^\eta_{b,T}\text{ }\phi^\eta(\textbf{k},\textbf{p})
\end{alignedat}
\label{eq:Wannier}
\end{equation}
with $H_{e,h}$ being the Hamiltonian for the electron-hole complex, $\varepsilon^\mu_{b,X}$ the exciton binding energy, and $\psi^\mu(\textbf{k})$ the exciton wave function with the compound index $\mu = n,\xi_h,\xi_e$  including the principal quantum number $n$ of the excitonic Rydberg-like series and the electron/hole valley indices $\xi_{e/h}$. In analogy, $H_{e_1,e_2,h}$ is the Hamiltonian for the electron-electron-hole complex, $\varepsilon^\eta_{b,T}$ the electron-electron-hole binding energy, and $\phi^\eta(\textbf{k},\textbf{p})$ the eigenfunction for the bound triplet with the index $\eta = \xi^{s_h}_h,\xi^{s_{e_1}}_{e_1},\xi^{s_{e_2}}_{e_2}$ specifying the valley and spin configuration of each constituent particle (see SM for more details). In the following, the single-particle valley index is implicitly assumed inside the $e/h$ indices in masses and energies.

We use the eigenfunctions $\psi^\mu(\textbf{k})$ and $\phi^\eta(\textbf{k},\textbf{p})$, obtained by solving the corresponding Wannier equations, via direct diagonalization for excitons \cite{berghauser2014analytical} and a variational approach for trions \cite{perea2024trion}, to transform into the exciton and trion basis, respectively. Here, we  introduce the exciton operators $X_\textbf{Q}^{\mu \dagger} = \sum_\textbf{k} \psi^{\mu*}(\textbf{k}) a^{\dagger}_{c,\textbf{k}+\tilde{m}_e \textbf{Q}}  a^{}_{v,\textbf{k}-\tilde{m}_h\textbf{Q}}$, where $\textbf{k}$ and $\textbf{Q}$ denote the relative and center-of-mass momentum, respectively, with the effective mass ratio $\tilde{m}_{e/h} = m_{e/h}/{(m_e+m_h)}$.
In analogy, we introduce the trion operators $T^{\eta \dagger}_\textbf{Q} = \sum_{\textbf{k}\textbf{p}} \phi^{\eta*}(\textbf{k},\textbf{p}) a^{\dagger}_{c_1,\alpha_{e_1}\textbf{Q+k}} a^{\dagger}_{c_2,\alpha_{e_2} \textbf{Q+p}} a^{}_{v,\textbf{p+k} - \alpha_h \textbf{Q}}$, which create a trion state consisting of two electrons ($e_1,e_2$) and a hole ($h$) \cite{perea2022trion,perea2024trion}. Here, $\textbf{Q}$ denotes the center-of-mass momentum of the trion, while $\textbf{k}$ and $\textbf{p}$ are the corresponding relative momenta, furthermore, $\alpha_i = m_i/( m_{e_1} + m_{e_2} + m_h)$ with $i \in [e_1,e_2,h]$, and $c_i$ denoting the conduction band of electron $e_i$.
The exciton and trion Hamilton operators are diagonal in their corresponding basis, i.e. $H_X = \sum_{\mu\textbf{Q}} E^{\mu}_{\textbf{Q},X} X^{\mu \dagger}_{\textbf{Q}}X^{\mu}_{\textbf{Q}}$ and $H_T = \sum_{\eta\textbf{Q}} E^{\eta}_{\textbf{Q},T} T^{\eta \dagger}_{\textbf{Q}}T^{\eta}_{\textbf{Q}}$ . Here,  the total exciton energy is given by $E^{\mu}_{\textbf{Q},X} = E^\mu_{g} + \varepsilon^\mu_{b,X} + \hbar^2\textbf{Q}^2/(2M^\mu_X)$ with the single-particle bandgap $E^\mu_{g} = E_c - E_v$ for the exciton state $\mu$, where $E_{c/v}$ denote the conduction/valence band energy. In analogy, the total trion energy is given by $ E^{\eta}_{\textbf{Q},T} = E_{c_1}+E_{c_2}-E_v + \varepsilon^\eta_{b,T} + \hbar^2\textbf{Q}^2/(2M^\eta_T)$. 
We will denote exciton and trion states as $X_\mu$ and $T_\eta$, respectively, with $\mu$ and $\eta$ labeling their valley and spin configuration, following the previously defined order of hole and electron indices.

In literature, the trion binding energy is often defined as the energy released after recombination of the bright exciton within the trion, a convention useful in photoluminescence experiments. As this definition is not suitable for ARPES, we define the trion binding energy as the total binding energy $\varepsilon^\eta_{b,T}$ obtained from the generalized Wannier equation. In addition, we introduce the electron–exciton binding energy $\Delta E^{T}_{e_i} = \varepsilon^\eta_{b,T} - \varepsilon^\mu_{b,X_{e_jh}}$ to denote the binding of a given electron $e_i$ to a specific exciton $X_{e_jh}$ within the trion state (as shown in the schematic picture, Fig.~\ref{fig:scheme}b) \cite{perea2022trion,perea2024electrically} . 

\begin{figure}[t!]
  \centering
  \includegraphics[width=\columnwidth]{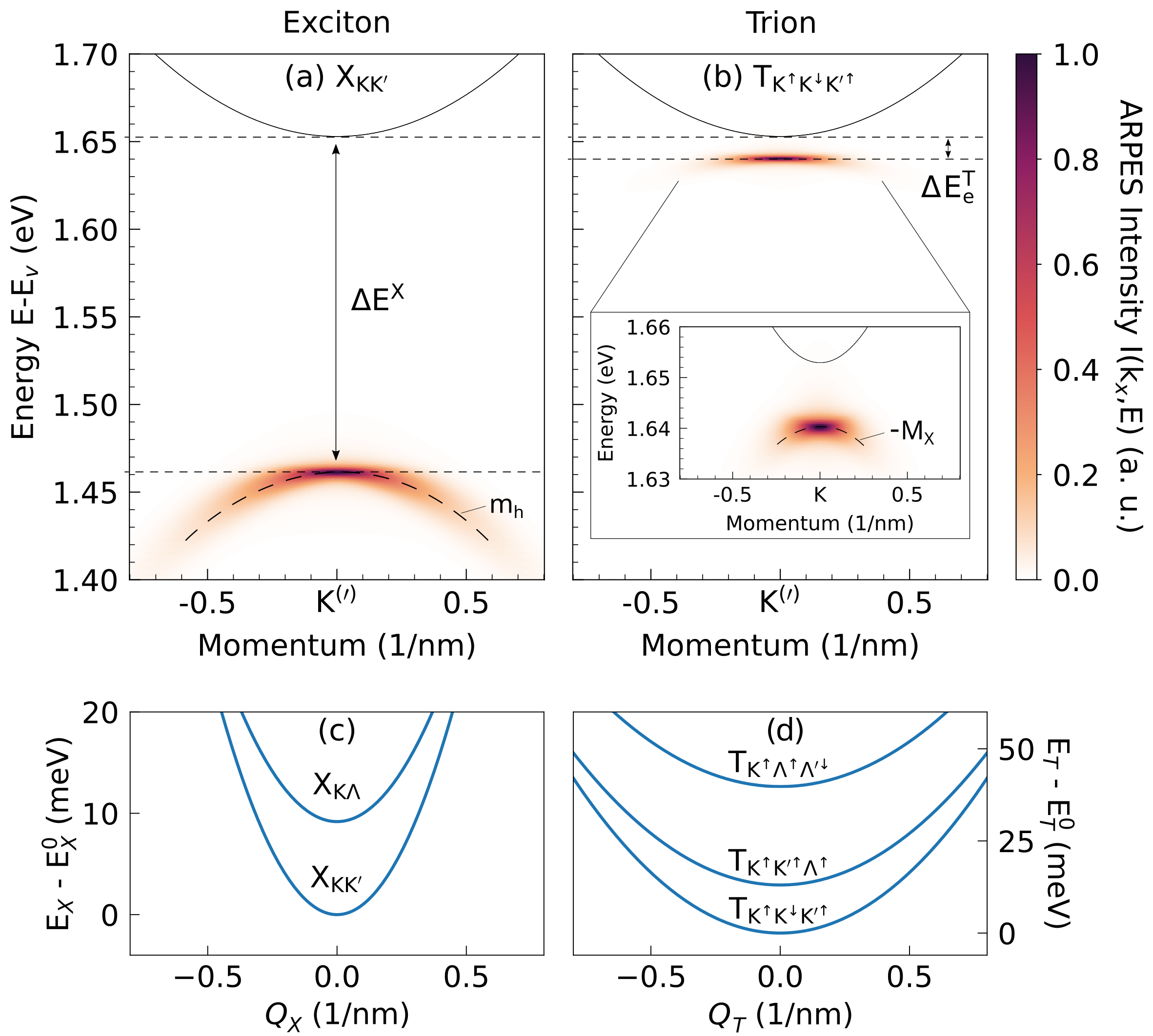}
  \caption{
ARPES intensity $I(k_x,E)$ for an n-doped WSe$_2$ monolayer along a momentum cut in the $k_x$ direction around the $K^{(\prime)}$ valley at $T=10$ K, shown as a function of energy for the lowest (a) exciton state X$_{\mathrm{KK}^{\prime}}$ and (b) trion state T$_{\mathrm{K}^{\uparrow}\mathrm{K}^{\downarrow}\mathrm{K}^{\prime\uparrow}}$. In both cases, the signal appears at the valley of the ejected electron, but its spectral position and shape differ. The exciton-related signal lies one exciton binding energy $\varepsilon^{}_{X,b}$ below the conduction-band minimum and follows the valence-band curvature (set by the hole effective mass $m_h$), whereas the trion signal is shifted by the much smaller electron–exciton binding energy $\Delta E^T_e$ and exhibits a nearly flat dispersion, reflecting the larger effective mass of the residual exciton ($M_X$). We take the valence-band maximum $\mathrm{E}_{v}$ as the reference energy level. (c,d) Energy landscape of the lowest-lying (c) excitonic and (d) trionic states as a function of their center-of-mass momentum $\mathbf{Q}_{X,T}$, with energies measured relative to the corresponding exciton or trion ground state $E^0_{X/T}$ at $\mathbf{Q}=0$ (for each valley configuration).
}
\label{fig:exciton_vs_trion}
\end{figure}

In this work, we focus on the low carrier-density and low-doping regime, where band renormalizations can be neglected and the few-body trion description provides an accurate representation of the system \cite{glazov2020optical}. 
In the case of excitons, we restrict to 1s states in the lowest spin-bright configuration and omit the spin index for simplicity.
Solving the Wannier-like equations for excitons and trions, Eq. \eqref{eq:Wannier}, we find that the lowest-lying excitonic states in WSe$_2$ monolayers are the momentum-dark $X_{\mathrm{KK}^\prime}$ and $X_{\mathrm{K\Lambda}}$ states \cite{malic2018dark, brem2020phonon}, while the lowest trionic states are T$_{\mathrm{K}^{\uparrow}\mathrm{K}^{\downarrow}\mathrm{K}^{\prime\uparrow}}$, T$_{\mathrm{K}^{\uparrow}\mathrm{K}^{\prime\uparrow}\mathrm{\Lambda}^{\uparrow}}$, and T$_{\mathrm{K}^{\uparrow}\mathrm{\Lambda}^{\uparrow}\mathrm{\Lambda}^{\prime\downarrow}}$ \cite{perea2024trion}, see Figs. \ref{fig:exciton_vs_trion}
c-d. In the absence of exchange interactions, these states are degenerate with respect to spin–valley configurations (e.g. K and K' valleys with opposite spin have the same energy), thus the signal is equal for degenerate spin-valley bands (e.g. T$_{\mathrm{K}^{\uparrow}\mathrm{K}^{\downarrow}\mathrm{K}^{\prime\uparrow}}$ is degenerate with T$_{\mathrm{K}^{\uparrow}\mathrm{K}^{\downarrow}\mathrm{K}^{\downarrow}}$). Electron–hole exchange is known to slightly lift this degeneracy both for excitons and trions \cite{yu2014dirac,druppel2017diversity,PhysRevB.96.085302,PhysRevB.105.085302,christianen2025asymmetrictrionsmonolayertransition}, but this typically small splitting is neglected here as our analysis focuses on the qualitative fingerprints of trions in ARPES spectra.
These states are used in the following as the initial configurations for computing the excitonic and trionic contributions to the ARPES signal. 

The corresponding initial (trion) and final (free electron+exciton) states for the ARPES process are written as $\ket{T_\eta}$ and $\ket{\textbf{k}}\otimes\ket{X_\mu}$ with $\textbf{k}$ being the wavevector of the emitted electron.
Inserting these states in Eq. (\ref{eq:arpes1}), we obtain the final equation for the ARPES signal of trion states 
\begin{equation}\label{eq:arpes2}
    \mathcal{I}({{\bf k}},h\nu) \propto \sum_{\eta \mu \textbf{Q} }\, \lvert  \mathcal{G}^{\eta\mu}_{{\bf Q} {\bf k}}\rvert^2 \, N^\eta_{{\bf Q}}  \,\delta\left(\Delta E^{\eta\mu}_{\textbf{Q}\textbf{k}}\right)
\end{equation}
where $\Delta E^{\eta\mu}_{\textbf{Q}\textbf{k}}=E^{}_{\textbf{k},e} + E^{\mu}_{\textbf{Q}-\textbf{k},X} - E^{\eta}_{{\bf Q},T} - h\nu$ with the free electron energy $E^{}_{\textbf{k},e}$ and the definition of exciton and trion energies given above. The photoemission signal depends directly on the trion occupation $N^\eta_{\bf Q}$ in the state $\eta$ and the trion center-of-mass momentum ${\bf Q}$, and the matrix element $\lvert  \mathcal{G}^{\eta\mu}_{{\bf Q} {\bf k}}\rvert^2$. The latter weights the contributions of different photoemission channels involving different electrons within the trion. These weights are determined by the internal structure of the trion wavefunction after tracing out the residual exciton degrees of freedom (see SM for more details).

\emph{ARPES signatures of trions at low temperatures}---%
The derived theoretical  approach is  applicable to a broad class of excitonic and trionic materials. Here, we focus on the exemplary case of n-doped WSe$_2$ monolayers at $T=10$ K and solve Eq.~\ref{eq:arpes2} to obtain the ARPES intensity $I(\mathbf{k},E)$ as a function of energy and momentum. At the cryogenic temperature considered, we focus only on the signature of the lowest momentum-dark  exciton $X_{\mathrm{KK}^\prime}$ and trion state $T_{\mathrm{K}^{\uparrow}\mathrm{K}^{\downarrow}\mathrm{K}^{\prime\uparrow}}$, the latter consisting of two electrons with the same mass.
Figure~\ref{fig:exciton_vs_trion} shows the ARPES intensity $I(k_x,E)$ along a momentum cut in the $k_x$ direction. Panel (a) illustrates the exciton signal, which reproduces well-known photoemission features \cite{rustagi2018photoemission, madeo2020directly,christiansen2019theory,meneghini2023hybrid}. The signal appears one exciton binding energy $\varepsilon^\mu_{X,b}$ below the conduction band minimum. At low temperatures, when the exciton distribution is sharply peaked around the center-of-mass momentum $\textbf{Q}=0$, the spectral shape follows the negative valence-band dispersion, in agreement with previous theoretical and experimental studies \cite{madeo2020directly, christiansen2019theory}. Panel (b) presents the trion ARPES signal. Compared to neutral excitons, the trion-related spectral feature is located much closer to the conduction-band minimum and shows a flatter momentum dispersion, deviating from the valence-band dispersion (see the inset).

Notably, while in photoluminescence trions appear approximately one electron--exciton binding energy $\Delta E^T_{e_i}$  below the exciton peak, in ARPES the trion signal lies instead higher in energy, as it is located one $\Delta E^T_{e_i}$ (where the index $e_i$ refers to the ejected electron) below the conduction band minimum. This difference reflects the fact that ARPES measures the energy required to remove a single electron from a bound complex. For an exciton, this corresponds to the exciton binding energy $\varepsilon^\mu_{b, X}$, whereas for a trion only the binding energy $\Delta E^T_{e_i}$ of the ejected electron to the remaining exciton matters. 
This  follows directly from energy and momentum conservation in Eq.~\ref{eq:arpes2}. At low temperatures, where $N^\eta_{\textbf{Q}}\simeq \delta_{\textbf{Q},0}$,  the delta function in Eq. \eqref{eq:arpes2} gives (using the previous definitions of $E^{\mu}_{\textbf{Q},X}$ and $E^{\eta}_{\textbf{Q},T}$)
\begin{equation}
E_{\textbf{k},e}-h\nu = E_{c_i} - |\Delta E^T_{e_i}| - \frac{\hbar^2 \textbf{k}^2}{(2M_X)},
\end{equation}
demonstrating that the trion signature appears one electron–exciton binding energy $\Delta E^T_{e_i}$ below the conduction-band minimum $E_{c_i}$ of the ejected electron $e_i$ (for $\textbf{k}=0$ corresponding to the center of the ejected electron valley, where the maximum of the signal appears).
The shape of the photoemission signal is governed by the relatively large effective exciton mass $M_X$, which  explains the flatter momentum dependence of the trion signal. The apparent slightly negative curvature  (see the inset in Fig. \ref{fig:exciton_vs_trion}b) arises from energy conservation in the photoemission process.

\emph{Double-peaked signal from mass-imbalanced trions}---%
So far, we have considered trion ARPES signals for n-doped WSe$_2$ monolayers under the assumption that only one trion state is occupied. In reality, the trion landscape in this material is more complex: several low-lying states contribute, and thermal occupation at finite temperatures can considerably modify ARPES spectra. In particular, a part of the low-energy response arises from mass-imbalanced trions composed of electrons from different valleys in the Brillouin zone exhibiting sizably distinct effective masses \cite{perea2024trion}. 
Because the internal structure of these mass-imbalanced trions is more complex, the corresponding photoemission process is expected to depend sensitively on the specific trion configuration. This leads to qualitatively distinct ARPES signatures schematically illustrated in  Figure~\ref{fig:scheme}b (right panel).
To account for this scenario, we consider a Boltzmann distribution at room temperature over the three energetically lowest trion states and analyze their combined photoemission signatures.  

\begin{figure}[t!]
  \centering
  \includegraphics[width=\columnwidth]{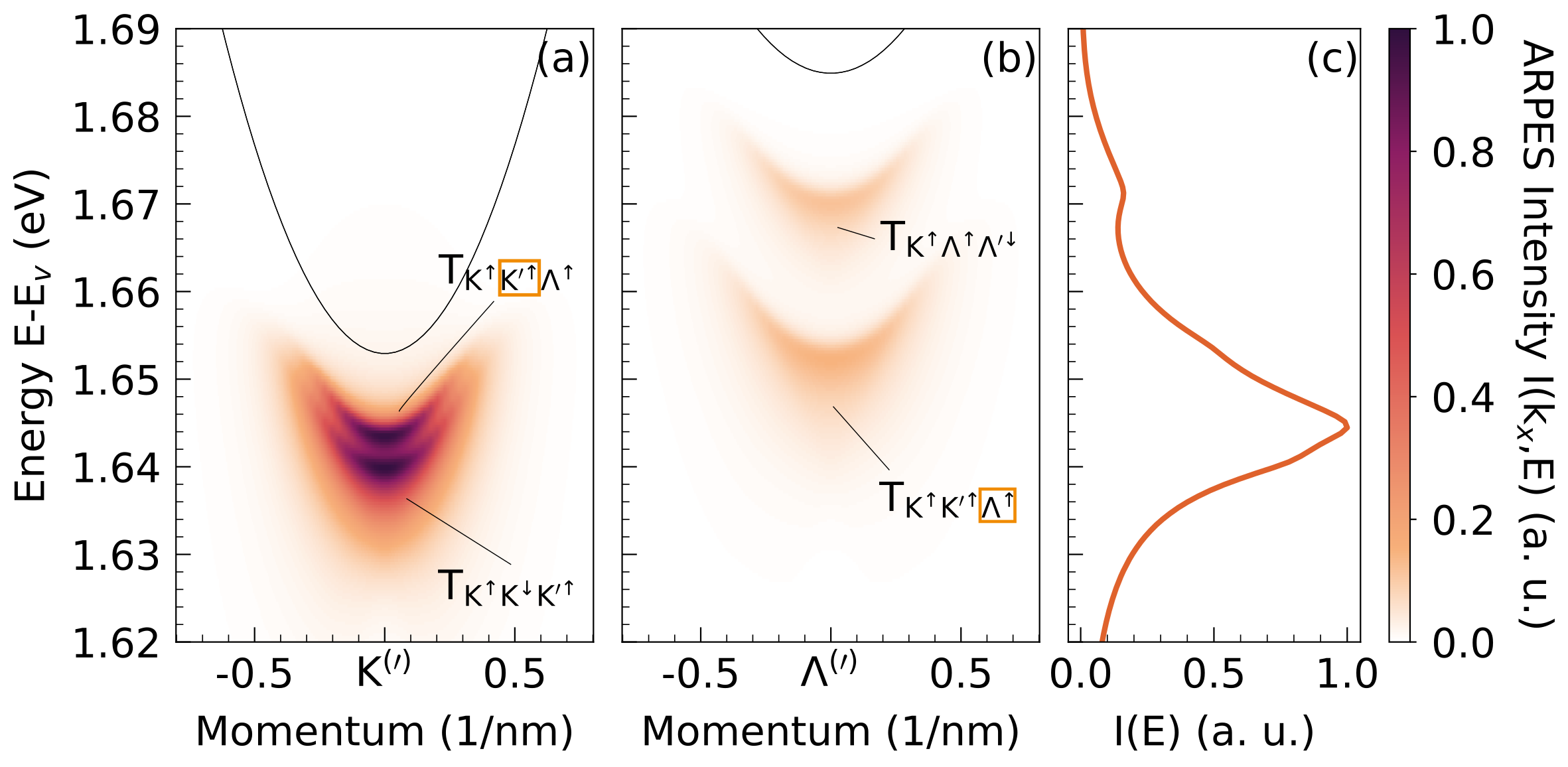}
  \caption{
Room temperature ARPES intensity for a n-doped WSe$_2$ monolayer considering a thermalized Boltzmann distribution of  the three energetically lowest trion states, T$_{\mathrm{K}^{\uparrow}\mathrm{K}^{\downarrow}\mathrm{K}^{\prime\uparrow}}$, T$_{\mathrm{K}^{\uparrow}\mathrm{K}^{\prime\uparrow}\mathrm{\Lambda}^{\uparrow}}$, and T$_{\mathrm{K}^{\uparrow}\mathrm{\Lambda}^{\uparrow}\mathrm{\Lambda}^{\prime\downarrow}}$.  Panels (a) and (b) show the signal around the K$^{(\prime)}$ and the $\Lambda^{(\prime)}$ valley, respectively, relative to the conduction-band minimum (solid black line).  For the mass-imbalanced trion T$_{\mathrm{K}^{\uparrow}\mathrm{K}^{\prime\uparrow}\mathrm{\Lambda}^{\uparrow}}$, the origin of the double signal is indicated by orange boxes around the corresponding ejected electron. (c) Momentum-integrated ARPES signal illustrating the emergence of multiple trionic resonances and the pronounced spectral broadening compared to the low-temperature case (Fig.~\ref{fig:exciton_vs_trion}), reflecting the broader center-of-mass momentum distribution at room temperature.
}
   \label{fig:trion_thermal}
\end{figure}

Figure~\ref{fig:trion_thermal} displays the valley-resolved  intensity $I(k_x,E)$ around the K$^\prime$ and the $\Lambda$ valley as well as the momentum-integrated ARPES intensity $I(E)$. For the mass-balanced trions T$_{{\mathrm{K}^{\uparrow}\mathrm{K}^{\downarrow}\mathrm{K}^{\prime \uparrow}}}$ and T$_{{\mathrm{K}^{\uparrow}\mathrm{\Lambda}^{\uparrow}\mathrm{\Lambda}^{\prime\downarrow}}}$, we find a single ARPES peak, respectively. The reason is that the two electrons within the trion are symmetry-equivalent: ejecting either one leaves the same residual exciton, and the two emission channels are degenerate  (see also Fig.~\ref{fig:scheme}b, left panel). The corresponding photoemission signals appear one electron-exciton binding energy below the conduction-band minimum at the momentum of the ejected electron, since $\Delta E^T_{e_1} = \Delta E^T_{e_2} \equiv \Delta E^T_e$. Quantitatively, the calculated electron–exciton binding energies are $\Delta E^T_e = 12\ \mathrm{meV}$ for T$_{\mathrm{K}^{\uparrow}\mathrm{K}^{\downarrow}\mathrm{K}^{\prime\uparrow}}$ and $\Delta E^T_e = 14\ \mathrm{meV}$ for T$_{{\mathrm{K}^{\uparrow}\mathrm{\Lambda}^{\uparrow}\mathrm{\Lambda}^{\prime\downarrow}}}$. The situation is richer for the mass-imbalanced trion T$_{{\mathrm{K}^{\uparrow}\mathrm{K}^{\prime\uparrow}\mathrm{\Lambda}^{\uparrow}}}$. Since the electrons at the K$^\prime$ and the $\Lambda$ valley have different effective masses (0.4 $\mathrm{m}_0$ and 0.6 $\mathrm{m}_0$ respectively, with $\mathrm{m}_0$ as the electron mass \cite{kormanyos2015k}), their binding to the residual exciton differs considerably. As a result, ejecting a K$^\prime$ electron or a $\Lambda$ electron leaves behind distinct excitonic configurations with different energies (see also Fig.~\ref{fig:scheme}b, right panel). This gives rise to a characteristic double-peak ARPES features: we predict trion signatures to appear at $\Delta E^T_\mathrm{K} = 8 \mathrm{meV}$ below the  conduction-band minimum at the K$^\prime$ valley  and $\Delta E^T_{\Lambda} = 31 \mathrm{meV}$ below the conduction-band minimum  at the $\Lambda$ valley, see Fig. \ref{fig:trion_thermal}. We highlight the two peaks originating from the same trion in Fig.~\ref{fig:trion_thermal} by enclosing the corresponding ejected electron label in an orange box.

The relative intensity of the two peaks is determined by the matrix element $\lvert \mathcal{G}^{\eta\mu}_{\bf Qk} \rvert^2$ in Eq. \eqref{eq:arpes2}, which contains the conditional probability of ejecting a given electron, obtained by tracing out the excitonic degrees of freedom from the trion wavefunction (see SM for more details). Because the excitonic components $X_{\mathrm{KK}^\prime}$ and $X_{\mathrm{K\Lambda}}$ possess different wavefunctions, $\lvert \mathcal{G}^{\eta\mu}_{{\bf Q}{\bf k}} \rvert^2$ differs for each excitonic configuration, resulting in distinct emission intensities for the two channels.
The momentum-integrated ARPES signal $I(E)$, shown in Fig.~\ref{fig:trion_thermal}c, reveals how the combined contributions of these resonances produce a richer and more intricate peak structure compared to the excitonic case. Furthermore, by comparing Fig.~\ref{fig:exciton_vs_trion} a-b and Fig.~\ref{fig:trion_thermal}, we find  that increasing the temperature from 10 K to room temperature markedly alters the trion signal: the features broaden both in energy and momentum, and the apparent curvature becomes positive due to the combined contributions of  excitonic and trionic dispersions, as reflected in the delta function of Eq.~\ref{eq:arpes2}. This behavior arises from the much broader center-of-mass momentum distribution of a thermal trion occupation.

\begin{figure}[t]
\centering
\includegraphics[width=\columnwidth]{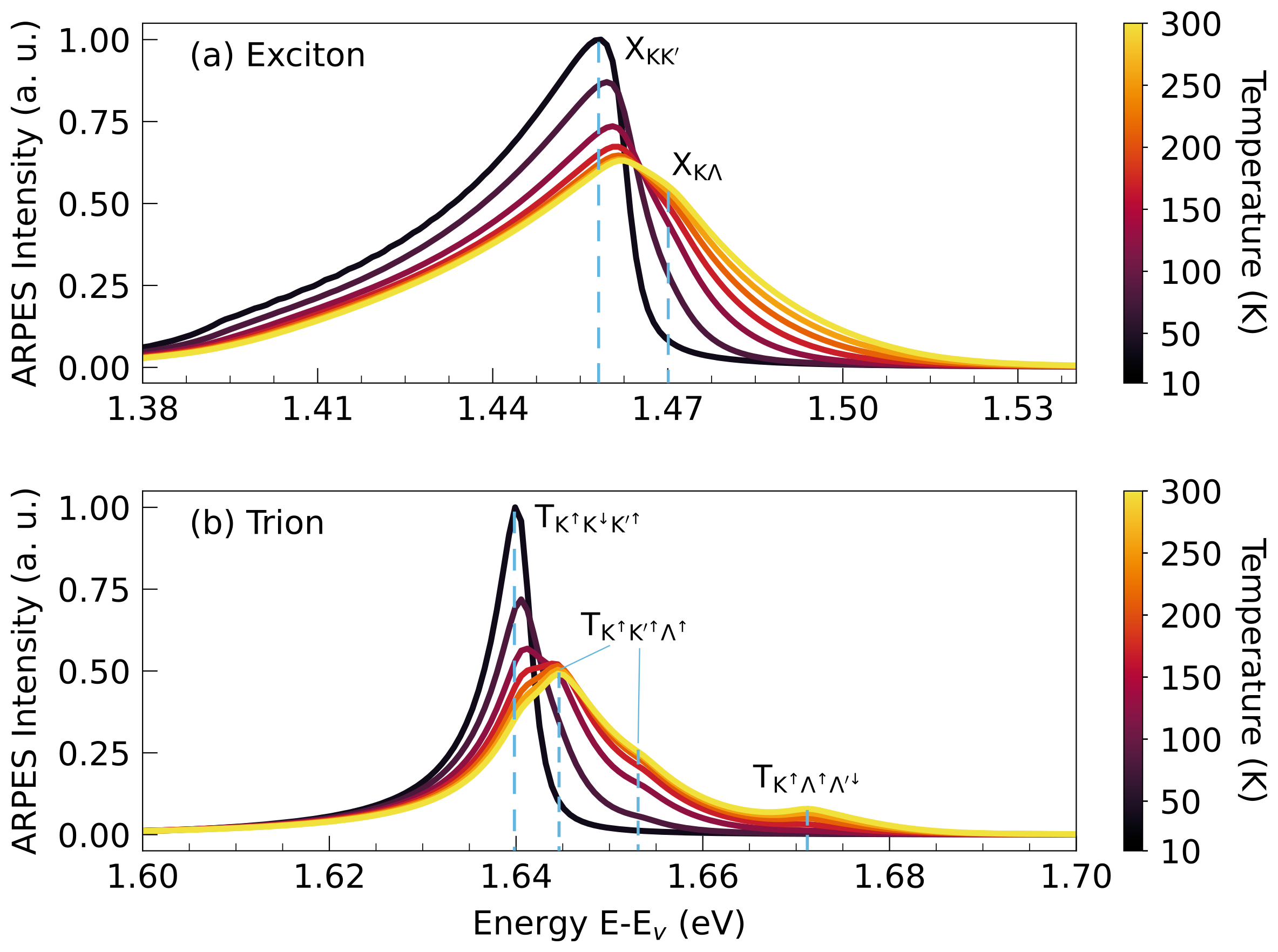}
\caption{
 Temperature dependence of the K and $\Lambda$ valley momentum-integrated ARPES signal $I(E)$ for a WSe$_2$ monolayer in the case of (a) excitons and (b) trions. For excitons, the low-energy states include X$_{\mathrm{KK}^{\prime}}$ and X$_{\mathrm{K}\Lambda}$, while for trions the relevant states are T$_{\mathrm{K}^{\uparrow}\mathrm{K}^{\downarrow}\mathrm{K}^{\prime\uparrow}}$, T$_{\mathrm{K}^{\uparrow}\mathrm{K}^{\prime\uparrow}\mathrm{\Lambda}^{\uparrow}}$, and T$_{\mathrm{K}^{\uparrow}\mathrm{\Lambda}^{\uparrow}\mathrm{\Lambda}^{\prime\downarrow}}$, see Figs. \ref{fig:exciton_vs_trion}c-d.
}
\label{fig:exc_trion_temperature}
\end{figure}

\emph{Temperature evolution of trion signatures}---%
We now examine the momentum-integrated temperature dependence of  ARPES spectra for n-doped WSe$_2$ monolayers and compare them with the excitonic response in the same material. 
Figure \ref{fig:exc_trion_temperature}a shows  the undoped case, where the exciton signal at low temperatures (dark curve) is broad and dominated by a single peak. A long asymmetric tail extends to lower energies, reminiscent of the “recoil effect” in optical spectra \cite{zipfel2022electron}. Such tails are intrinsic and arise from the joint constraints of momentum and energy conservation (delta function in Eq. \ref{eq:arpes2}). As the temperature increases, the spectrum broadens asymmetrically towards higher energies and a second peak emerges. These two features (indicated by dashed blue lines) originate from the two lowest excitonic resonances of WSe$_2$, namely X$_{\mathrm{KK}^\prime}$ and X$_{\mathrm{K\Lambda}}$, see Fig. \ref{fig:exciton_vs_trion}c. Since the two excitonic states are separated by approximately 10 meV, a sufficiently high temperature (around 100 K) is required to thermally populate the higher X$_{\mathrm{K\Lambda}}$ state and make its contribution visible in ARPES.

Figure~\ref{fig:exc_trion_temperature}b illustrates the trion case in n-doped WSe$_2$ monolayers, which exhibits a richer temperature-dependent evolution compared to the excitonic signal. At low temperatures, the spectrum is dominated by the energetically lowest  mass-balanced trion state T$_{\mathrm{K}^{\uparrow}\mathrm{K}^{\downarrow}\mathrm{K}^{\prime\uparrow}}$, resulting in a single peak. As the temperature increases, the thermal population of higher-energy trions produces a multiplet structure: four distinct subpeaks appear within a 50 meV window (highlighted by dashed blue lines and labels). They arise from the three energetically lowest trion states, T$_{\mathrm{K}^{\uparrow}\mathrm{K}^{\downarrow}\mathrm{K}^{\prime\uparrow}}$, T$_{\mathrm{K}^{\uparrow}\mathrm{K}^{\prime\uparrow}\mathrm{\Lambda}^{\uparrow}}$, and T$_{\mathrm{K}^{\uparrow}\mathrm{\Lambda}^{\uparrow}\mathrm{\Lambda}^{\prime\downarrow}}$, lying close in energy (see Fig. \ref{fig:exciton_vs_trion}d). Here,  the mass-imbalanced trion  T$_{\mathrm{K}^{\uparrow}\mathrm{K}^{\prime\uparrow}\mathrm{\Lambda}^{\uparrow}}$ contributes with a double-peak, as discussed in Fig. \ref{fig:trion_thermal}.
Similar to the exciton case, a thermal activation is required to bridge the energy gaps between the trion states. At low temperatures, only the lowest trion states contributes significantly. Around 50 K, the second trion  T$_{\mathrm{K}^{\uparrow}\mathrm{K}^{\prime\uparrow}\mathrm{\Lambda}^{\uparrow}}$ ($\approx$13 meV above the lowest state), begins to contribute. Only at higher temperatures ($\approx$100–150 K) the third trion state, T$_{\mathrm{K}^{\uparrow}\mathrm{\Lambda}^{\uparrow}\mathrm{\Lambda}^{\prime\downarrow}}$ ($\approx$40 meV above the lowest state) also becomes important and results in a clearly observable additional peak. Given that the energy separation between these features is on the order of tens of meV, resolving them could be challenging with the energy resolution currently achievable in femtoseconds ARPES experiments \cite{schmitt2022formation,bange2023ultrafast,bange2024probing}. However, the characteristic double-peak structure arising from the mass-imbalanced trion appears at separate valleys and might thus be accessible in experiments.

\emph{Conclusion}---%
We have developed a material-specific and predictive microscopic framework for describing excitonic and trionic signatures in ARPES spectra of doped atomically-thin semiconductors. By explicitly computing the momentum-resolved response of the exemplary, n-doped WSe$_2$ monolayer, we answer  the still unresolved question of how charged excitons manifest in ARPES. We demonstrate that trions generate robust fingerprints including specific shifts in energy and a modified spectral shape with respect to excitonic signatures. In particular, we predict a distinctive double-peak feature for mass-imbalanced trions and a thermally activated multiplet of resonances, providing quantitative criteria for their unambiguous detection in ARPES experiments.  The developed approach  establishes a general framework for studying many-body Coulomb complexes in doped two-dimensional semiconductors in ARPES measurements.

\bibliographystyle{apsrev4-1}
\bibliography{references}

\section{Acknowledgements}
 We acknowledge financial support from the Deutsche Forschungsgemeinschaft  (DFG) through  the priority program 535247173 SPP 2244 and the regular project 542873285. Calculations  were conducted on the Lichtenberg high-performance computer of the TU Darmstadt.

 \clearpage
\pagestyle{empty}

\foreach \p in {1,...,4}{   
  \begin{figure}[p]
    \centering
    \includegraphics[page=\p,width=\textwidth]{supplemental_information.pdf}
  \end{figure}
}

\end{document}